\documentclass[pdflatex,sn-mathphys-num]{sn-jnl}

\usepackage{tikz}
\usetikzlibrary{quantikz}
\usepackage{graphicx}
\usepackage{multirow}
\usepackage{amsmath,amssymb,amsfonts}
\usepackage{amsthm}
\usepackage{mathrsfs}
\usepackage[title]{appendix}
\usepackage{xcolor}
\usepackage{textcomp}
\usepackage{manyfoot}
\usepackage{booktabs}
\usepackage{algorithm}
\usepackage{algorithmicx}
\usepackage{algpseudocode}
\usepackage{listings}
\usepackage{caption}

\theoremstyle{thmstyleone}%
\theoremstyle{thmstyletwo}%

\theoremstyle{thmstylethree}%

\begin{document}

\title[Quantum Port]{Quantum Port: Gamification of quantum teleportation for public engagement}

\author*[1,2]{\fnm{Pak Shen} \sur{Choong}} \email{pakshenchoong@mmu.edu.my}

\author[3]{\fnm{Aqilah} \sur{Rasat}} \email{aqilahrasat@gmail.com}
\equalcont{These authors contributed equally to this work.}

\author[4]{\fnm{Afiqa} \sur{Nik Aimi}} \email{nomnomyush@gmail.com}
\equalcont{These authors contributed equally to this work.}

\author[2,5]{\fnm{Nurisya} \sur{Mohd Shah}} \email{risya@upm.edu.my}
\equalcont{These authors contributed equally to this work.}

\affil[1]{Faculty of Computing \& Informatics, Multimedia University, Persiaran Multimedia, 63100 Cyberjaya, Selangor, Malaysia}

\affil[2]{Institute for Mathematical Research (INSPEM), Universiti Putra Malaysia (UPM), 43400 UPM Serdang, Selangor, Malaysia}

\affil[3]{AR Display Resources, Selangor, Malaysia}

\affil[4]{SEGi College Subang Jaya, Menara A, Edumetro, Persiaran Subang Permai, USJ 1, 47500 Subang Jaya, Selangor, Malaysia}

\affil[5]{Department of Physics, Faculty of Science, Universiti Putra Malaysia (UPM), 43400 UPM Serdang, Selangor, Malaysia}

\abstract{Concepts on quantum physics are generally difficult for the general public to understand and grasp due to its counter-intuitive nature and requirement for higher level of mathematical literacy. With categorical quantum mechanics (CQM), quantum theory is re-formalized into a more intuitive diagrammatic approach, which we will refer to as the first level of transformation, to improve the accessibility and readability of quantum theory to a broader audience since the mathematical details are embedded into diagrammatic rules. Taking inspiration from this diagrammatic approach, we propose the second level of transformation by gamifying the diagrammatic rules of quantum teleportation into a quantum card game called Quantum Port. In this work, we discuss the gamification of quantum teleportation and provide a moderator guideline to use Quantum Port as a public engagement or learning module.}


\maketitle

\section{Introduction} \label{SecIntroduction}

Since the inception of quantum theory through wave and matrix formalisms by Schr\"{o}dinger \cite{Schrodinger2003} and Heisenberg \cite{Heisenberg2015}, independently and respectively, quantum theory has contributed to the fundamental understanding of our physical world and will continue to be an integral part of modern technologies through the second quantum revolution \cite{Dowling2003}. On 7 June 2024, the United Nations (UN) proclaimed 2025 to be the International Year of Quantum Science and Technology (IYQ 2025) to celebrate and recognize the importance of quantum theory to humanity \cite{IYQ2025}, particularly with the recent breakthroughs of quantum computing \cite{Supremacy2019,Willow2024} which may help to realize the possibility of using quantum machine learning and artificial intelligence for the benefit of humanity at large \cite{Biamonte2017,Cerezo2022,GESDA2024,QuantumDelta2024}.

However, quantum theory remains a difficult subject for the general public to understand, especially for those without formal training in physics and mathematics. This is due to the counter-intuitive nature of fundamental concepts in quantum theory, such as quantum entanglement \cite{Einstein1935} and quantum contextuality \cite{Bell1966,KochenSpecker1968}. With the recent announcement on the Willow quantum chip that demonstrated the possibility of real-time quantum error correcting \cite{Willow2024}, we may soon enter the age of fault-tolerant quantum computing \cite{Preskill2025}. Hence, it is important to improve quantum literacy among the general public so that quantum theory becomes more accessible to non-specialists. In order to achieve this goal, various teaching and learning modules \cite{Violaris2023,Dundar-Coecke2023,Sang2024,Patino2024,Sanz2024,Goorney2024,1Levy2025} and quantum games \cite{Goff2006,Kultima2021,Piispanen2023,Evenbly2024,Gaunkar2024,Piispanen2025,Armbruster2025,2Levy2025} have been proposed to engage audiences of varying levels of backgrounds, so that the quantum ecosystem will be more inclusive than just quantum scientists, researchers, and technology companies. In this paper, we describe a quantum card game based on quantum teleportation, called Quantum Port, that features two levels of transformation. The first level of transformation comes from the diagrammatic approach to quantum theory \cite{Coecke2017}, whereas the second level of transformation comes from the gamification of the diagrammatic rules to quantum teleportation.

Our paper will be divided into the following sections. In Section \ref{SecMathQuanTele}, we will briefly describe the mathematical formalism and the protocol of quantum teleportation. In Section \ref{SecFirstTransformation}, we will explain how a picture of quantum teleportation is derived in the diagrammatic approach of categorical quantum mechanics. In Section \ref{SecSecondTransformation}, we will introduce the Quantum Port card game and discuss the gamification process. In Section \ref{secEngagementModule}, we will provide a moderator guideline to introduce the game and debrief after the gaming session. We conclude our work with some future prospects for Quantum Port.

\section{Quantum teleportation} \label{SecMathQuanTele}

Quantum teleportation \cite{Nielsen2010} begins with two parties, Alice and Bob, sharing a pair of entangled qubits in one of the Bell bases,
\begin{align}
\ket{\Phi}_{AB}^+ & = \frac{1}{\sqrt{2}} (\ket{00} + \ket{11}), \\
\ket{\Phi}_{AB}^- & = \frac{1}{\sqrt{2}} (\ket{00} - \ket{11}), \\
\ket{\Psi}_{AB}^+ & = \frac{1}{\sqrt{2}} (\ket{01} + \ket{10}), \\
\ket{\Psi}_{AB}^- & = \frac{1}{\sqrt{2}} (\ket{01} - \ket{10}).
\end{align}
Without loss of generality, we focus on the Bell basis $\ket{\Phi}_{AB}^+$ in our discussion. The other Bell bases follow the same mathematical arguments, and the Pauli correction required for each of the Bell bases will be summarized in Table \ref{TablePauliCorrection}.

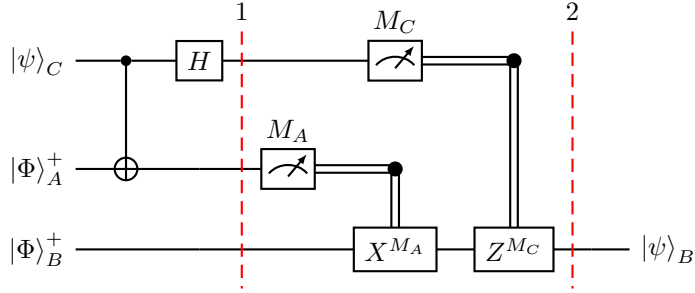
\begin{figure}[h]
\centering
\begin{quantikz}
\lstick{$\ket{\psi}_C$}   & \ctrl{1} & \gate{H} \slice{1} & \qw           & \meter{$M_C$}  & \cwbend{2} \\
\lstick{$\ket{\Phi}_A^+$} & \targ{}  & \qw                & \meter{$M_A$} & \cwbend{1} \\
\lstick{$\ket{\Phi}_B^+$} & \qw      & \qw                & \qw           & \gate{X^{M_A}} & \gate{Z^{M_C}} \slice{2} & \qw & \rstick{$\ket{\psi}_B$} \qw
\end{quantikz}
\caption{Quantum teleportation} \label{FigureQuantumTeleportation}
\end{figure}

Alice has a third qubit, $\ket{\psi}_C = \alpha \ket{0} + \beta \ket{1}$ that she wants to send to Bob. Instead of sending the qubit $\ket{\psi}_C$ directly to Bob, she can utilize the Bell pair and the assistance of classical communication to achieve the same task in a more secure way. The protocol for quantum teleportation is illustrated in Figure \ref{FigureQuantumTeleportation}.

Initially, the shared Bell state between Alice and Bob, and the qubit that Alice wants to send to Bob is given as:
\begin{align}
\ket{\psi}_{CAB} & = (\alpha \ket{0}_C + \beta \ket{1}_C) \otimes \frac{1}{\sqrt{2}} (\ket{00}_{AB} + \ket{11}_{AB}) \nonumber \\
& = \frac{1}{\sqrt{2}} (\alpha \ket{000} + \alpha \ket{011} + \beta \ket{100} + \beta \ket{111}).
\end{align}
After the CNOT gate, the above quantum state becomes:
\begin{align}\label{qteleport-eqn}
\ket{\psi}_{CAB} = \frac{1}{\sqrt{2}} (\alpha \ket{000} + \alpha \ket{011} + \beta \ket{110} + \beta \ket{101}).
\end{align}
After the Hadamard gate, the quantum state at Step 1 in Figure \ref{FigureQuantumTeleportation} will become:
\begin{align}
\ket{\psi}_{CAB} & = \frac{1}{\sqrt{2}} \left[ \frac{\alpha}{\sqrt{2}} (\ket{0}_C + \ket{1}_C) \otimes \ket{00}_{AB} + \frac{\alpha}{\sqrt{2}} (\ket{0}_C + \ket{1}_C) \otimes \ket{11}_{AB} \right. \nonumber \\ & \phantom{\frac{1}{\sqrt{2}}} \qquad \left. + \frac{\beta}{\sqrt{2}} (\ket{0}_C - \ket{1}_C) \otimes \ket{10}_{AB} + \frac{\beta}{\sqrt{2}} (\ket{0}_C - \ket{1}_C) \otimes \ket{01}_{AB} \right] \nonumber \\
& = \frac{1}{2} \left[ \alpha \ket{000} + \alpha \ket{100} + \alpha \ket{011} + \alpha \ket{111} + \beta \ket{010} - \beta \ket{110} + \beta \ket{001} - \beta \ket{101} \right] \nonumber \\
& = \frac{1}{2} \left[ \ket{00}_{CA} (\alpha \ket{0}_B + \beta \ket{1}_B) + \ket{01}_{CA} (\beta \ket{0}_B + \alpha \ket{1}_B) + \ket{10}_{CA} (\alpha \ket{0}_B - \beta \ket{1}_B) \right. \nonumber \\ & \phantom{\frac{1}{2}} \qquad \left. + \ket{11}_{CA} ( -\beta \ket{0}_B + \alpha \ket{1}_B) \right].
\end{align}

In Step 2, quantum measurements will be performed on Alice's qubits (qubit $A$ and $C$) and the appropriate Pauli correction will be applied to Bob's qubit (qubit $B$) to recover the quantum information that is teleported. In short,
\begin{enumerate}
\item If the quantum measurement results in $\ket{00}_{CA}$, nothing ($\sigma_4 = \left( \begin{smallmatrix} 1 & 0 \\ 0 & 1 \end{smallmatrix} \right)$) will be performed;
\item If the quantum measurement results in $\ket{01}_{CA}$, Pauli $X$, $\sigma_1 = \left( \begin{smallmatrix} 0 & 1 \\ 1 & 0 \end{smallmatrix} \right)$ will be performed;
\item If the quantum measurement results in $\ket{10}_{CA}$, Pauli $Z$, $\sigma_3 = \left( \begin{smallmatrix} 1 & 0 \\ 0 & -1 \end{smallmatrix} \right)$ will be performed;
\item If the quantum measurement results in $\ket{11}_{CA}$, Pauli $Y$ up to a global phase of $i$, $i \sigma_2 = \left( \begin{smallmatrix} 0 & 1 \\ -1 & 0 \end{smallmatrix} \right)$ will be performed.
\end{enumerate}

One may repeat the same calculation as above for other Bell bases and obtain the results shown in Table \ref{TablePauliCorrection}.
\begin{table}[h]
\captionsetup{width=0.9\textwidth}
\caption{Pauli correction (in term of $\sigma_i,\, i=1,2,3,4$) up to a global phase equivalence on Bob's qubit to obtain the correct quantum information} \label{TablePauliCorrection}
\begin{tabular}{|c|c|c|c|c|}
\hline
& 00 & 01 & 10 & 11 \\ \hline
$\ket{\Psi}_{AB}^+$ & $\sigma_1$ & $\sigma_4$ & $\sigma_2$ & $\sigma_3$ \\ \hline
$\ket{\Psi}_{AB}^-$ & $\sigma_2$ & $\sigma_3$ & $\sigma_1$ & $\sigma_4$ \\ \hline
$\ket{\Phi}_{AB}^+$ & $\sigma_4$ & $\sigma_1$ & $\sigma_3$ & $\sigma_2$ \\ \hline
$\ket{\Phi}_{AB}^-$ & $\sigma_3$ & $\sigma_2$ & $\sigma_4$ & $\sigma_1$ \\ \hline
\end{tabular}
\end{table}

\section{First level of transformation: Diagrammatic approach to quantum processes} \label{SecFirstTransformation}

Categorical Quantum Mechanics (CQM) is a diagrammatic approach to quantum theory that was introduced by Abramsky and Coecke \cite{Abramsky2004}. The motivation for CQM is to find a generalized framework for quantum theory, with category theory as its mathematical backbone, and refine it until we can find a complete description of quantum mechanics. In the years since the introduction of CQM, Coecke and his many collaborators have refined CQM to include descriptions of mixed states and purification \cite{Selinger2007,Coecke2008Mixed,Coecke2012}, quantum measurement and observables \cite{Coecke2008Measure,Coecke2011,Coecke2013}, quantum and classical channels \cite{Coecke2012,Coecke2014}, and causal structures \cite{Coecke2014Causal, Kissinger2019}. This section follows closely the derivation of quantum teleportation in the highly accessible textbook by Bob Coecke and Aleks Kissinger on CQM \cite{Coecke2017}.   

The basic building blocks of CQM are processes, depicted as a trapezoid with an incoming wire and outgoing wire. In this section, we adopt the bottom-to-top convention of reading a diagram. So, the incoming wire is at the bottom and the outgoing wire is at the top. These wires represent finite-dimensional quantum systems. 

\begin{figure}[h!]
\centering
\includegraphics[scale=0.2]{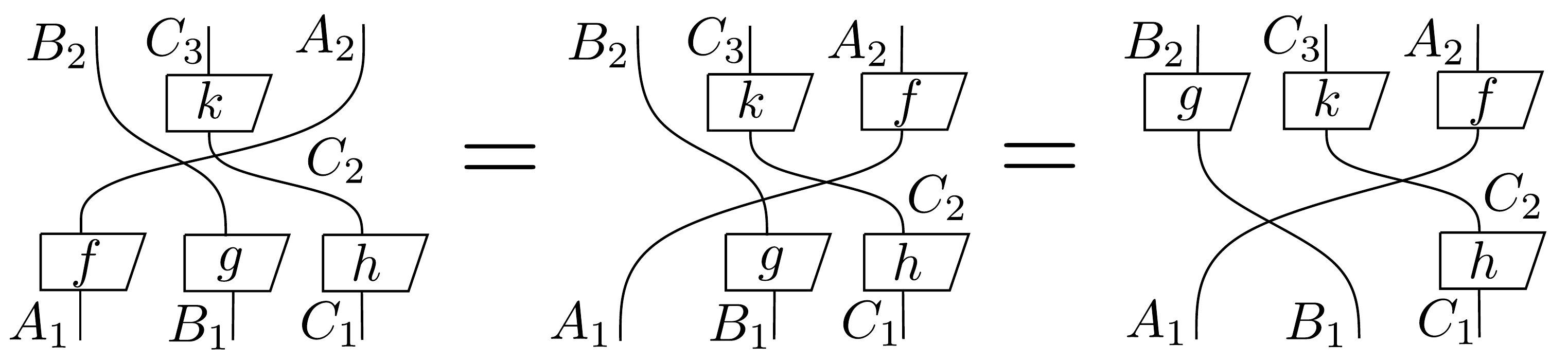}
\caption{A composition of processes and its equivalent diagrams} \label{compose-process-ex}
\end{figure}

Processes obey certain rules that allow for the box in a process to move along its wires. The composition of processes is itself a process and can be represented as multiple diagrams, as depicted in Figure \ref{compose-process-ex}. Figure \ref{compose-process-ex} also shows that processes can be composed sequentially and in parallel. 

\begin{figure}[h!]
\begin{minipage}[t]{.49\textwidth}
\centering
\includegraphics[scale=0.2]{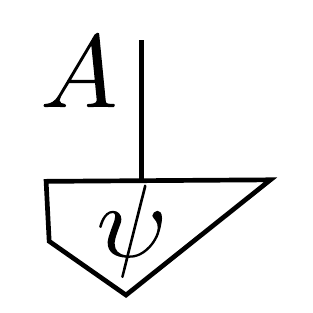}
\caption{State $\psi$ of system $A$} \label{state}
\end{minipage}
\hfill
\begin{minipage}[t]{.49\textwidth}
\centering
\includegraphics[scale=0.2]{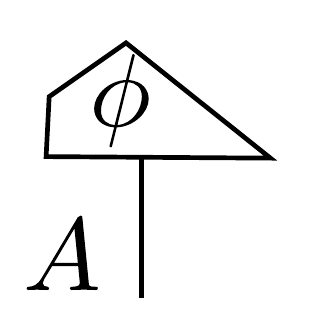}
\caption{Effect $\phi$ of system $A$} \label{effect}
\end{minipage}
\end{figure}

A state $\psi$ of a system $A$ is a process without an input wire (see Figure \ref{state}). In contrast, an effect $\phi$ of a system $A$ is a process without an output wire (see Figure \ref{effect}). 

\begin{figure}[h!]
\begin{minipage}[t]{.49\textwidth}
\centering
\includegraphics[scale=0.2]{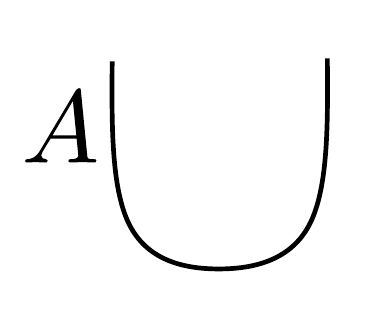}
\caption{Cup for system $A$} \label{cup}
\end{minipage}
\hfill
\begin{minipage}[t]{.49\textwidth}
\centering
\includegraphics[scale=0.2]{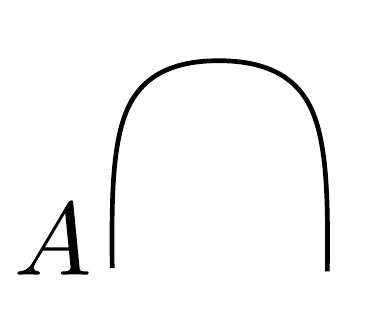}
\caption{Cap for system $A$} \label{cap}
\end{minipage}
\end{figure}

CQM treats entanglement as fundamental to quantum theory. It is required for every system, say $A$, to have a non-separable bipartite state, called cup (see Figure \ref{cup}), and a non-separable bipartite effect, called cap (see Figure \ref{cap}), which satisfy Equation \ref{eq-yank-cup-cap}.   

\begin{equation}
\includegraphics[scale=0.2]{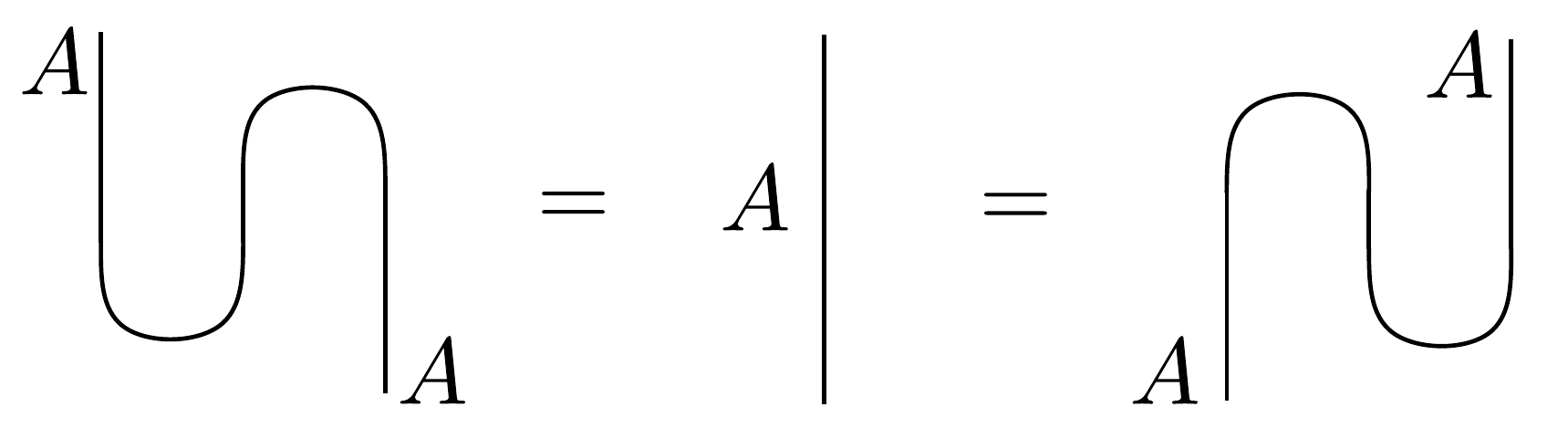}
\label{eq-yank-cup-cap}
\end{equation}

\begin{figure}[h!]
\begin{minipage}[t]{.49\textwidth}
\centering
\includegraphics[scale=0.2]{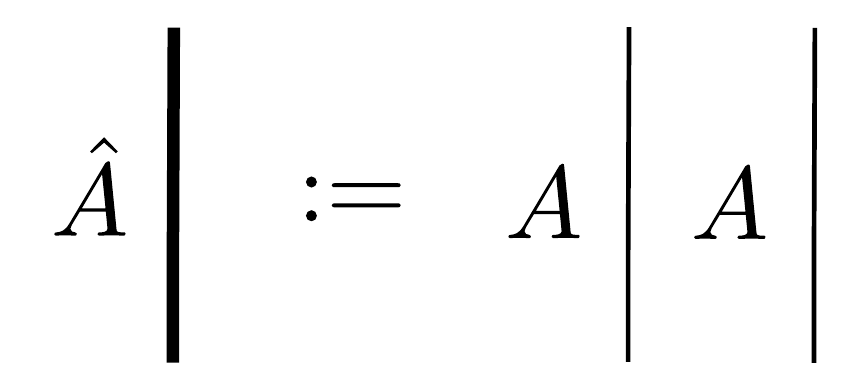}
\caption{Double wire for system $A$} \label{double-wire}
\end{minipage}
\hfill
\begin{minipage}[t]{.49\textwidth}
\centering
\includegraphics[scale=0.2]{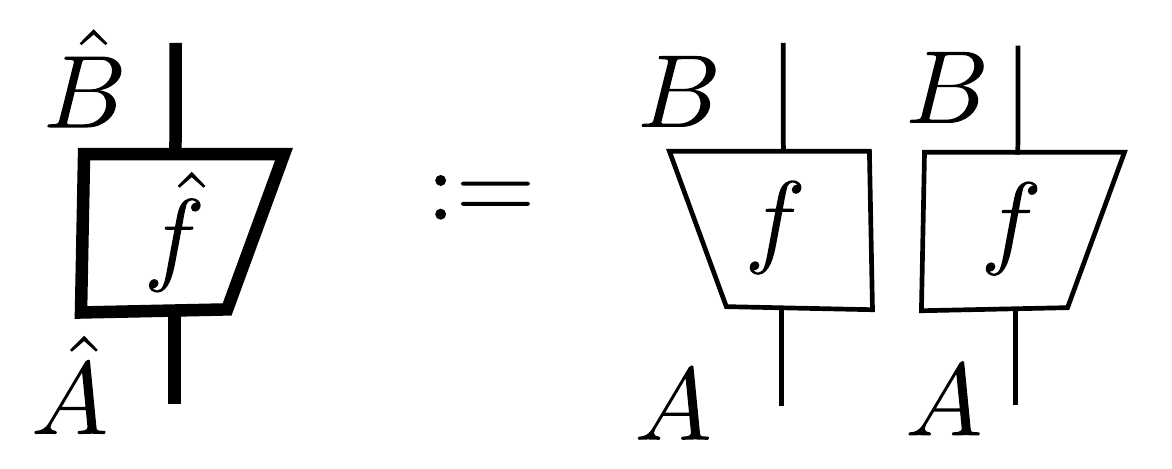}
\caption{Double process for process $f$} \label{double-process}
\end{minipage}
\end{figure}

To simplify a diagram -- so that we do not need to draw a process twice in a single diagram -- we can apply the ``doubling" convention introduced by Coecke. The doubling of a wire is shown in Figure \ref{double-wire} and the doubling of a process is shown in Figure \ref{double-process}. The doubling convention allows for the inclusion of quantum states and operations which are not necessarily pure. These are the processes in the ``double world" which do not have a single counterpart, i.e. they cannot be redrawn as the (separable) parallel composition of a single process and its conjugate.

Henceforth, we shall use the ``doubling" convention and denote them by their single counterpart with the hat symbol. For those double diagrams that may not have a single counterpart, we shall label them without the hat symbol. Furthermore, we shall forego labeling the wires henceforth. Unless otherwise stated, each diagram will involve only systems of the same dimension. That is, a diagram will not have a system with dimension $d_1$ \textbf{and} a system with dimension $d_2$ where $d_1\neq d_2$. At this point, we have introduced the diagrams needed to provide a picture of quantum teleportation.

\begin{figure}[h!]
\centering
\includegraphics[scale=0.2]{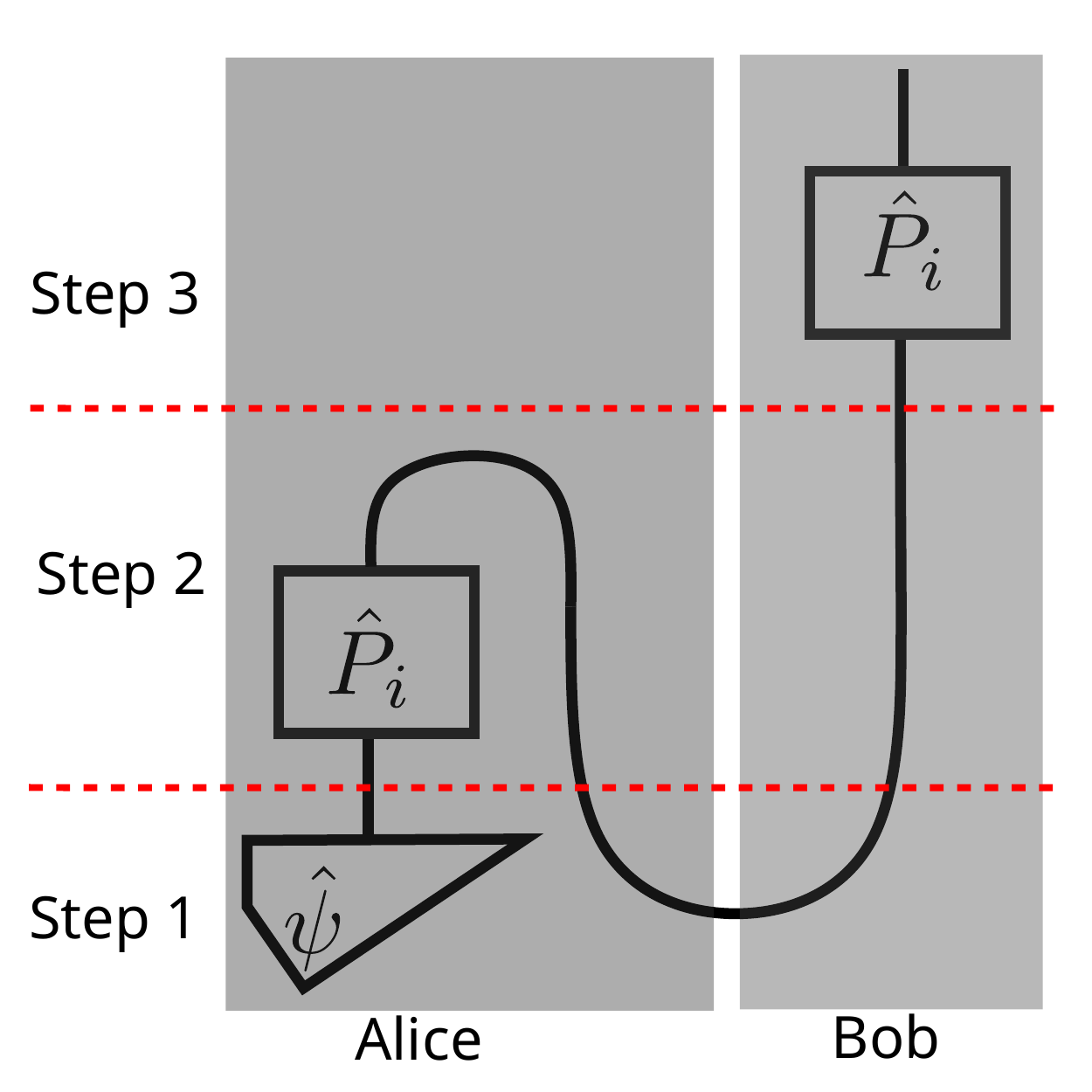}
\caption{A first picture of quantum teleportation} \label{qteleport-cqm}
\end{figure}

In Figure \ref{qteleport-cqm}, the wires in the diagram are qubits. The state $\hat{\psi}$ is the state that Alice needs to send to Bob. The cup in Step 1 is the shared Bell state between Alice and Bob. The cap in Step 2 is the measurement Alice performs on her part of the Bell state and $\hat{\psi}$, where the process $\hat{P}_i$ is the Pauli matrix that represents the measurement outcome obtained by Alice. Finally, in Step 3 on Bob's side, Bob applies the Pauli correction needed to obtain $\hat{\psi}$, so it must be equal to the Pauli matrix in Step 2. We can simplify the diagram for quantum teleportation in the following way:
\begin{equation}
\includegraphics[scale=0.2]{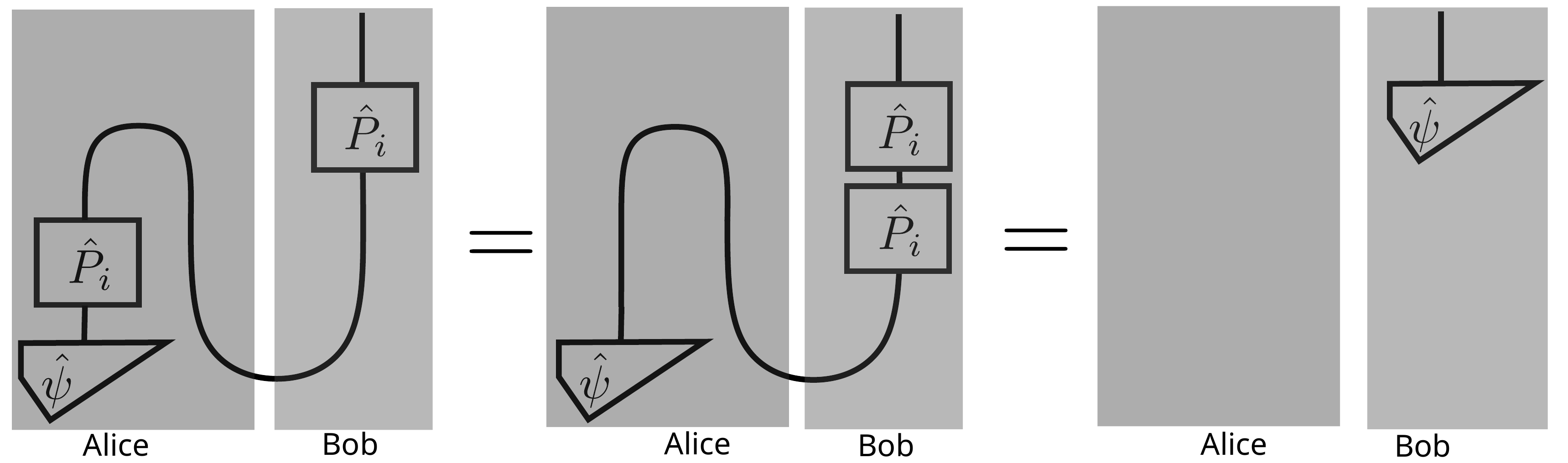}
\label{eq-qt-pauli}
\end{equation}
In Equation (\ref{eq-qt-pauli}), the $\hat{P}_i$ on Alice’s side is slid along the wire to Bob’s side in the first equality, and the second equality is due to Equation \ref{eq-yank-cup-cap} and the self-unitarity of the Pauli matrix in the diagram. 

\begin{figure}[h!]
\centering
\includegraphics[scale=0.2]{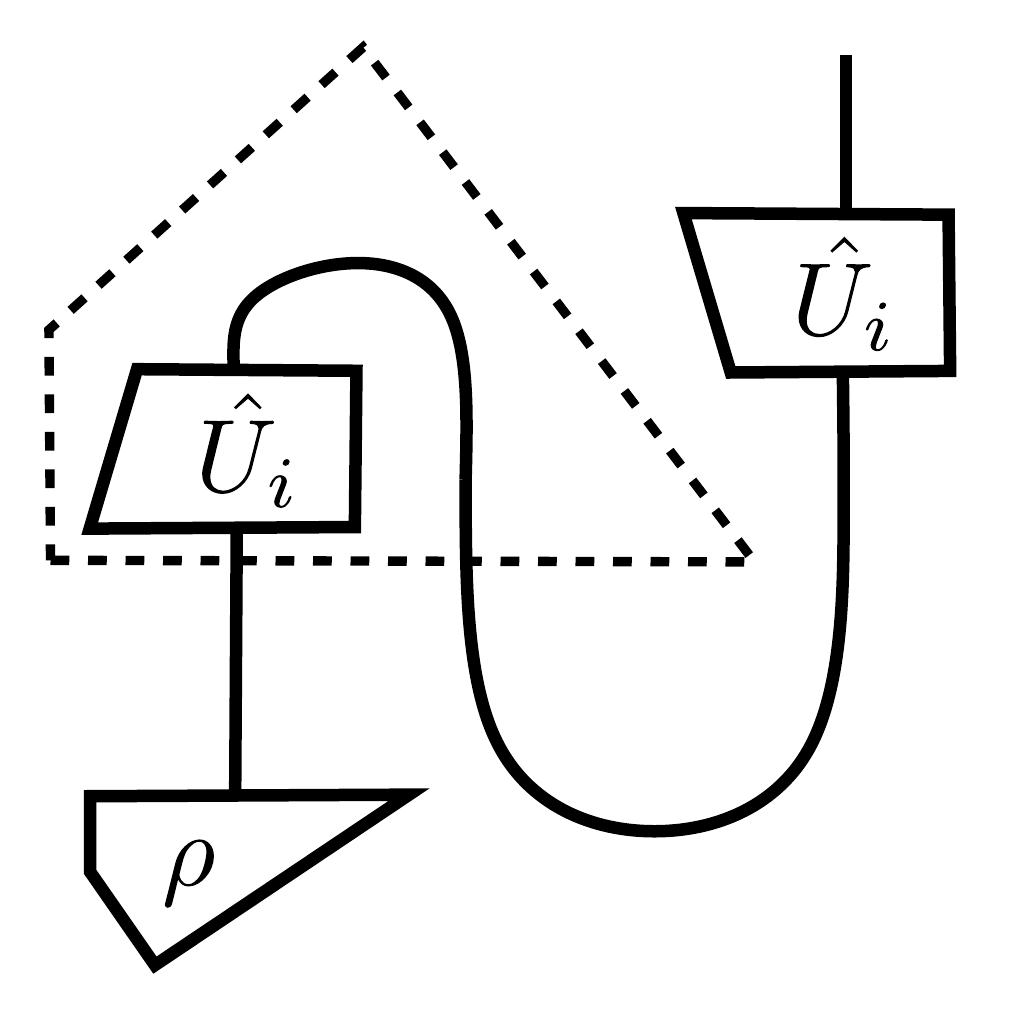}
\caption{The general post-measurement diagram for quantum teleportation} \label{qt-unitary}
\end{figure}

The picture of quantum teleportation in Figure \ref{qteleport-cqm} can be generalized so that the wires represent (double) qu$d$its with $d\geq 2$. Then, the set of $\hat{P}_i$ can be replaced with a set of unitaries $\hat{U}_i$, while $\hat{\psi}$ can be replaced with a state not necesarily pure, say $\rho$. This is shown in Figure \ref{qt-unitary}, where the unitaries are indexed to represent the outcomes of the measurement performed by Alice, and the dashed line encompasses this measurement.

The diagram in Figure \ref{qt-unitary} is a post-measurement picture of quantum teleportation. That is, there is an implicit assumption that Alice communicates the outcome of her measurement to Bob by the matching index $i$ of the unitaries. If Alice does not communicate her measurement’s outcome, the state $\rho$ cannot be transported to Bob but instead, Bob receives a maximally mixed state. Alice \textbf{needs} to communicate the outcome of her measurement to Bob and this communication is classical.

While the doubling convention is certainly convenient, there is a structural reason behind its introduction, namely, to distinguish quantum systems from classical systems, where quantum systems are double wires while classical systems are single wires. A classical system is a system encoded with classical values via a fixed orthonormal basis. Indeed, when we do an experiment in a quantum system, the results that we observe from the experiment are recorded as classical data – information that we can read. 

\begin{figure}[h!]
\begin{minipage}[t]{.49\textwidth}
\centering
\includegraphics[scale=0.2]{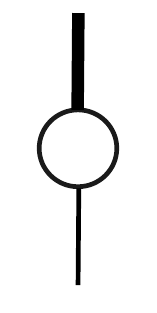}
\caption{The encode process, which has a single wire as its input and a double wire as its output} \label{encode}
\end{minipage}
\hfill
\begin{minipage}[t]{.49\textwidth}
\centering
\includegraphics[scale=0.2]{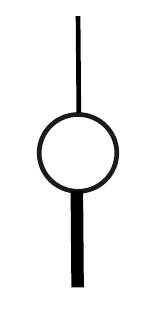}
\caption{The measure process which has a double wire as its input and a single wire as its output} \label{measure}
\end{minipage}
\end{figure}

Again, we take the example of quantum teleportation, where the possible outcomes of Alice’s measurement are given by an orthonormal basis. The interaction between quantum and classical systems is depicted by two processes: Encode, which has a single wire as its input and a double wire as its output, as shown in Figure \ref{encode}; and measure, which has a double wire as its input and a single wire as its output, as shown in Figure \ref{measure}. Their names give a hint on how they are interpreted in quantum theory, i.e. encode is the process of encoding classical data into a quantum system, whereas measure is the process of extracting classical data from a quantum system.

\begin{figure}[h!]
\centering
\includegraphics[scale=0.2]{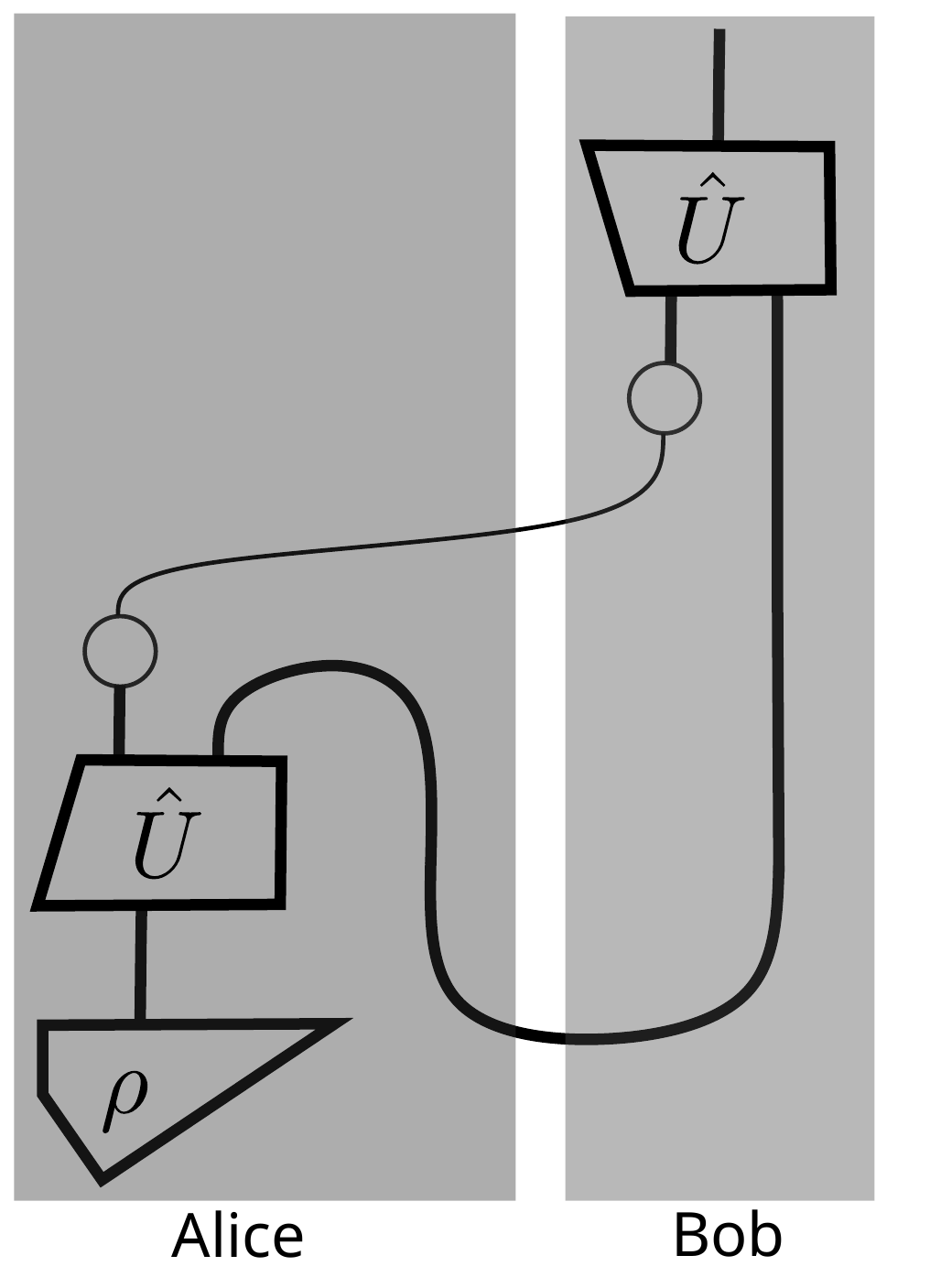}
\caption{A picture of quantum teleportation with classical communication} \label{qt-complete}
\end{figure}

Now that we have a way to describe a quantum-classical interaction, we can complete our picture of quantum teleportation, which includes a classical communication between Alice and Bob, as shown in Figure \ref{qt-complete}. The unitary process in the diagram has become a controlled unitary as its application by Bob depends on the outcome of the measurement performed by Alice. That is, Alice extracts classical data through measurement on the state of the bipartite system on her side. She then sends the aforementioned data to Bob and Bob encodes the classical data into the quantum system on his side. We can rewrite the diagram in a similar manner to Equation \ref{eq-qt-pauli} to show that the quantum state $\rho$ is indeed transported successfully from Alice to Bob. 

\section{Second level of transformation: Gamification} \label{SecSecondTransformation}

The second level of transformation comes from the gamification of the diagrammatic rules to quantum teleportation. Before we begin, we would like to clarify that all in-game terminologies will be capitalized and in the \texttt{typewriter} font.

\subsection{Introduction to in-game components and terminologies}

Quantum Port is a tabletop game played by two players. At the beginning of the game, each player is given a deck of 41 cards, consisting of \texttt{Circuit Cards} and \texttt{Action Cards}; 4 \texttt{Quantum Data Coins} (\texttt{QDC}), 4 \texttt{Classical Data Coins} (\texttt{CDC}), and a \texttt{Classical Coin}. In addition to the cards and coins, a \texttt{Four-Sided Die} should be included in a complete set.

The \texttt{Circuit Cards} are further divided into the \texttt{Classical Circuits} (illustrated with thin, white lines), \texttt{Quantum Circuits} (illustrated with thick, black lines), \texttt{Swap Circuits} (a mixture of \texttt{Classical} and/or \texttt{Quantum Circuits}), \texttt{Entangler}, and \texttt{Pauli Correction}. Meanwhile, there are four types of \texttt{Action Cards}, namely \texttt{Information Jammer}, \texttt{Circuit Destroyer}, \texttt{Information Revealer}, and \texttt{Eavesdropper}.

The \texttt{Data Coins} have symbols on them. A \texttt{QDC} has its tail (also known as coin back or reverse) ``Q" and head (also known as coin front or obverse) being one of the following symbols: ``$\Psi^+$", ``$\Psi^-$", ``$\Phi^+$", and ``$\Phi^-$". Meanwhile, a \texttt{CDC} has its tail ``C" and head being one of the following symbols: ``00", ``01", ``10", and ``11". A \texttt{Classical Coin} has ``C" on both its head and tail.

\subsection{Setups and winning conditions}

To set up the game, each player performs the following procedures:
\begin{enumerate}
\item Choose some \texttt{QDC}: 1 \texttt{QDC} for the \texttt{Beginner} setup, 2 \texttt{QDC} for the \texttt{Default} setup;
\item Shuffle their deck of cards and pick 4 from the top of the deck;
\item Place the chosen \texttt{QDC} on player's left with the head facing down, along with their \texttt{Classical Coin} in a vertical alignment (refer to Figures \ref{QPEasy} and \ref{QPHard} for \texttt{Beginner} and \texttt{Default} setups respectively). The locations of \texttt{QDC} and \texttt{Classical Coin} determine how to build the \texttt{Quantum Teleportation Circuit} (\texttt{QTC}) later on.
\end{enumerate}

In the \texttt{Beginner} setup, the fastest player to obtain a \texttt{QDC} wins the game. In the \texttt{Default} setup, the player who holds the most \texttt{QDC} on hand at the end of the game will win. There are two ways for a player to obtain a \texttt{QDC}:
\begin{enumerate}
\item[\texttt{Aim 1}] Complete a \texttt{QTC} and execute an instance of quantum teleportation;
\item[\texttt{Aim 2}] Steal a \texttt{QDC} while the opponent attempts to execute an instance of quantum teleportation.
\end{enumerate}

\subsection{Eligible \texttt{Turns} and \texttt{Moves}} \label{subsecTurnandMove}

Each player take \texttt{Turns} to perform \texttt{Moves} in order to obtain a \texttt{QDC}. In each \texttt{Turn}, a player is limited to \textbf{two} \texttt{Moves}. A \texttt{Move} consists of either one of the following:
\begin{enumerate}
\item Use a \texttt{Circuit Card} to build their \texttt{QTC};
\item Discard a \texttt{Circuit Card} in their \texttt{QTC};
\item Draw a card from their deck;
\item Use an \texttt{Action Card} to sabotage their opponent;
\item Roll the \texttt{Four-Sided Die} to decide the corresponding \texttt{CDC} to be advanced;
\item Advance a \texttt{CDC} one \texttt{Classical Circuit Card} forward.
\end{enumerate}

At the end of their \texttt{Turn}, the player must discard any extra cards so that they only have \textbf{a maximum of four cards on hand}.

\subsection{Building a \texttt{Quantum Teleportation Circuit}}

In this section, we describe how a player can obtain a \texttt{QDC} through \texttt{Aim 1}. In short, there are two parts to \texttt{Aim 1}. The player must first build a \texttt{QTC}, and only after they have completed it, they can execute an instance of quantum teleportation. The building blocks of a \texttt{QTC} are the \texttt{Circuit Cards}. Each \texttt{Circuit Card} bears a diagram which represents a circuit element needed to build a \texttt{QTC}.

As explained in Section \ref{SecFirstTransformation}, the protocol for quantum teleportation has two parts: The quantum part consists of a shared entangled state, a measurement, and an error correction; The classical part consists of communication where the transmitting party tells the receiving party about the outcome of the measurement in the quantum part. The design of \texttt{Circuit Cards} tries to reflect this as much as possible. 

\texttt{Quantum Circuit Cards} consist of \texttt{Entangler}, \texttt{Pauli Correction}, and \texttt{Quantum Circuits}, whereas \texttt{Classical Circuit Cards} consist of only the \texttt{Classical Circuits}. There exists \texttt{Swap Circuits}, where as its name implies, swap between the positions of classical and/or quantum ``wires". Here, ``wire" is used in the same context as in Section \ref{SecFirstTransformation}.

The quantum part of a \texttt{QTC} must begin with an \texttt{Entangler} and end with a \texttt{Pauli Correction}. \texttt{Quantum} or \texttt{Swap Circuits} can be used in between these two cards without any limits, at the player's discretion. Meanwhile, the classical part of a \texttt{QTC} must consist of the same amount of cards as the quantum part.

\begin{figure}[h!]
\centering
\includegraphics[scale=0.35]{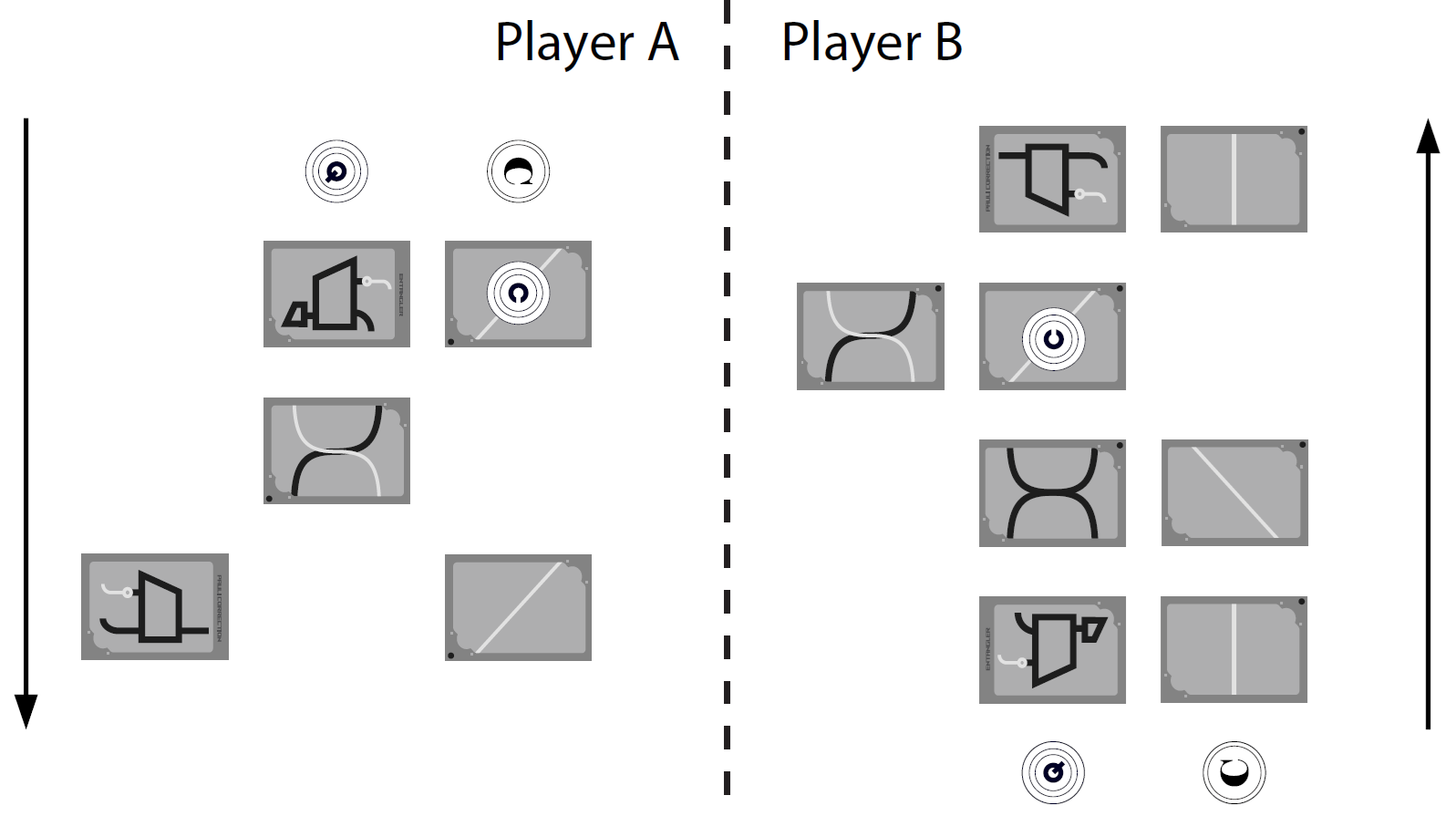}
\caption{\texttt{Beginner} setup of Quantum Port. Both players have completed their \texttt{QTC} and are advancing their \texttt{CDC} respectively. Player A requires three \texttt{Moves} to execute an instance of quantum teleportation, while Player B requires two \texttt{Moves}.} \label{QPEasy}
\end{figure}

Figure \ref{QPEasy} shows a screenshot of the \texttt{Beginner} setup gameplay, where both players have completed their own \texttt{QTC}. We note that the cards are oriented sideways to reflect the orientation of the cards in a practical game. The \texttt{QDC} and \texttt{Classical Coin} are always on the player's left and the \texttt{QTC} is read from left to right, so the arrows show how to read the \texttt{QTC} from both player's perspective. 

The quantum part of Player A’s \texttt{QTC} is a typical three-card build with \texttt{Entangler}, \texttt{Swap Circuit}, and \texttt{Pauli Correction}. \texttt{QDC} marks the beginning of the quantum part of \texttt{QTC}, hence \texttt{Entangler} is placed on the immediate right of Player A’s \texttt{QDC}. The \texttt{Quantum Circuit} on the \texttt{Swap Circuit} goes from top to bottom, so \texttt{Pauli Correction} must be placed below the \texttt{Swap Circuit}. The classical part of Player A’s \texttt{QTC} also follows a three-card build that begins from the \texttt{Classical Coin}, and simply needs to be connected by following the orientations of the \texttt{Classical Circuits}.

Meanwhile, Player B's \texttt{QTC} shows a four-card build. Notice that both of the \texttt{Swap Circuits} have a ``redundant" wire. It is understood that Player B chooses the necessary \texttt{Quantum Circuits} on both \texttt{Swap Circuits} to complete the quantum part of \texttt{QTC}.

\subsection{Executing quantum teleportation}

Once a player has completed a \texttt{QTC}, they can execute an instance of quantum teleportation. A \texttt{CDC} is selected by rolling a \texttt{Four-Sided Die}, where the outcome of the die roll is revealed to both players. The player then refers to the \texttt{Dice Rule} in Table \ref{TableDiceRule} and identify the corresponding \texttt{CDC} based on the outcome of the die roll and the \texttt{QDC} selected at the beginning of the game. The \texttt{CDC} is placed on the first \texttt{Classical Circuit Card} of player's \texttt{QTC} with its head facing down. This whole process is considered as one \texttt{Move}. Afterwards, advancing the \texttt{CDC} one \texttt{Classical Circuit Card} forward will take one \texttt{Move}. An instance of quantum teleportation is completed when the \texttt{CDC} advances beyond the final \texttt{Classical Circuit Card} in the \texttt{QTC} and the player obtains the \texttt{QDC}. As an example, in Figure \ref{QPEasy}, both players are attempting to advance their \texttt{CDC} to complete a single instance of quantum teleportation.
\begin{table}[h]
\caption{\texttt{Dice Rule} to decide the corresponding \texttt{CDC}} \label{TableDiceRule}

\begin{tabular}{|c|c|c|c|c|}
\hline
& 00 & 01 & 10 & 11 \\ \hline
$\Psi^+$ & 1 & 4 & 2 & 3 \\ \hline
$\Psi^-$ & 2 & 3 & 1 & 4 \\ \hline
$\Phi^+$ & 4 & 1 & 3 & 2 \\ \hline
$\Phi^-$ & 3 & 2 & 4 & 1 \\ \hline
\end{tabular}
\end{table}

\begin{figure}[h!]
\centering
\includegraphics[scale=0.35]{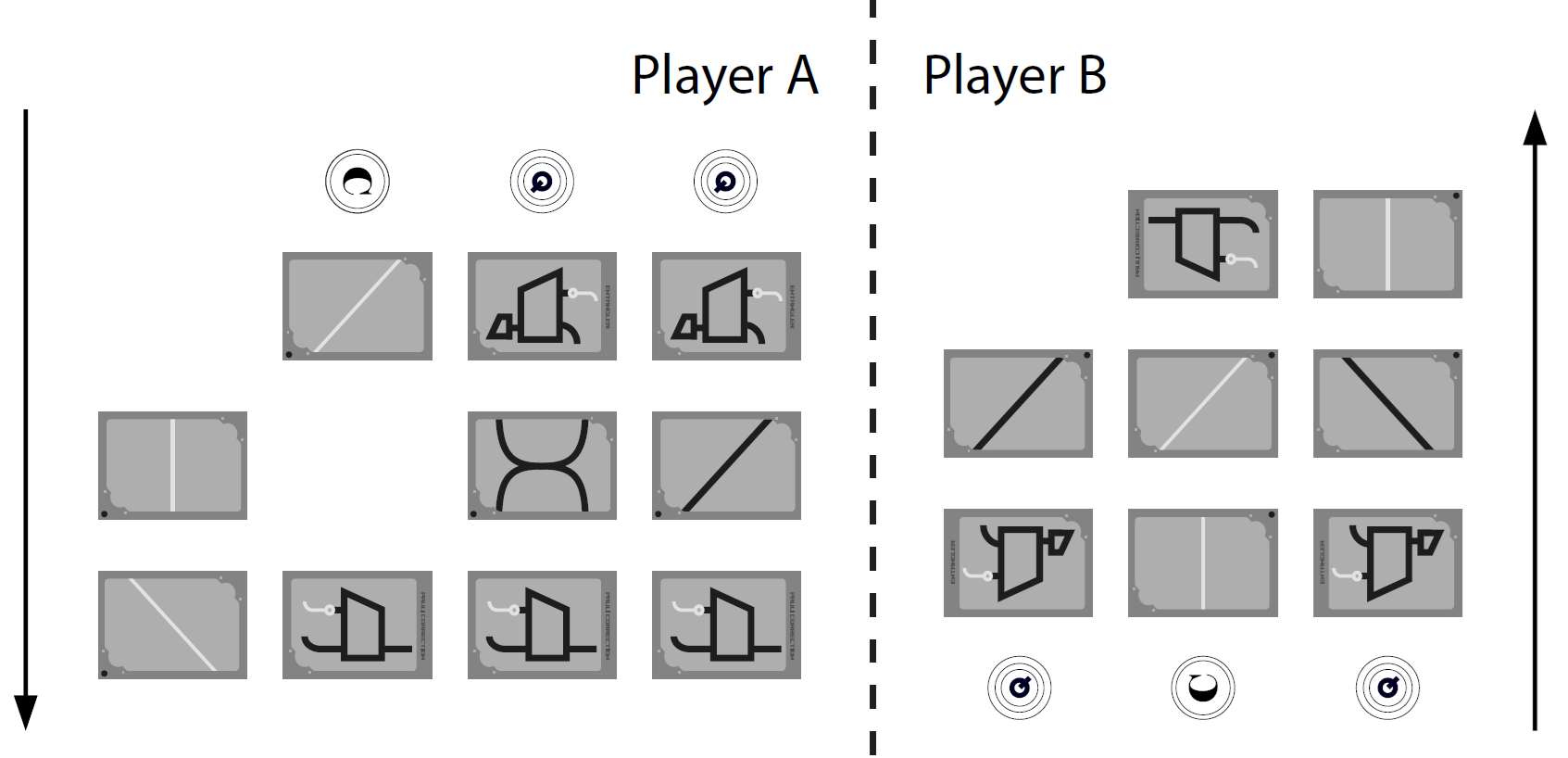}
\caption{\texttt{Default} setup of Quantum Port. Both players have completed their \texttt{QTC}.} \label{QPHard}
\end{figure}

Figure \ref{QPHard} shows a screenshot of the \texttt{Default} setup of gameplay with two \texttt{QDC}. Each \texttt{QDC} has to begin with their respective \texttt{Entangler}, however it is possible to have one \texttt{Pauli Correction} for two \texttt{QTC}, as shown by Player B's build. On the other hand, Player A in Figure \ref{QPHard} has three \texttt{Pauli Correction}, where one of the \texttt{QDC} has two choices of \texttt{Quantum Circuits} to execute an instance of quantum teleportation. The \texttt{Swap Circuit} does not play the role of swapping the position of two processes. Instead, it plays the role of branching the \texttt{Quantum Circuits} into 2 possible ``paths".

Both scenarios depicted in Figure \ref{QPHard} are in conflict with CQM's compositional rules of wires and a process. We partially remedy this issue by prohibiting the execution of concurrent quantum teleportation. This means that one must advance a \texttt{CDC} beyond the final \texttt{Classical Circuit Card} in the \texttt{QTC} to complete an instance of quantum teleportation before beginning the next instance of quantum teleportation. While this does not completely resolve the conflict with CQM, we stress that the quantum and classical parts of an instance of quantum teleportation is well-defined at any moment during the gameplay.

\subsection{Opponent sabotage}

A player can sabotage their opponent by using an \texttt{Action Card} during their \texttt{Turn}. There are four types of \texttt{Action Cards}:

\begin{enumerate}
\item \texttt{Circuit Destroyer}, which destroys one \texttt{Circuit Card} in the opponent’s \texttt{QTC};
\item \texttt{Information Jammer}, which freezes the opponent’s \texttt{CDC} from advancing for one \texttt{Turn};
\item \texttt{Information Revealer}, which reveals the opponent's advancing \texttt{CDC};
\item \texttt{Eavesdropper}, which steals the opponent's \texttt{QDC} if the player correctly guesses the symbol on the \texttt{QDC}. If the guess is wrong, the player is prohibited from stealing the \texttt{QDC} for the rest of the game.
\end{enumerate}

To obtain a \texttt{QDC} through \texttt{Aim 2}, player can either use \texttt{Eavesdropper} by itself or \texttt{Information Revealer} followed by \texttt{Eavesdropper}. Ideally, the latter will give the player a sure chance of correctly guessing the \texttt{QDC}, while the former has a 1 in 4 chance of guessing correctly. For the \texttt{Default} setup, \texttt{Information Revealer} only shows the corresponding \texttt{CDC} to one of the two possible \texttt{QDC}. The player still needs to guess the correct \texttt{QDC} that the opponent is trying to teleport.

\section{Quantum Port as a public engagement or learning module} \label{secEngagementModule}

Since the game is designed to match diagrammatically, it is possible to avoid using jargons when teaching the game. In addition, since most of the physical operations performed in quantum teleportation are embedded into the gaming rules, players learn about quantum teleportation without actively realizing it. The game was first tested as a prototype during an engagement event, called \textit{Not Art, Not Quantum}, held at the Telekom Museum, Malaysia on 16 November 2024. Subsequently, it was featured at the \textit{Quantum Art Valentine's Symposium 2025} on 14 February 2025 at Naresuan University, Thailand and \textit{ASEAN Quantum Summit 2025} on 10-12 December 2025, Universiti Teknologi Malaysia, Malaysia. Based on our experiences, we provide a moderator guideline and the necessary details to be included in the post-game debrief session so that Quantum Port can be run either as a public engagement or learning module.

\subsection{Moderator guideline}

We suggest that the moderator should first understand quantum teleportation from Sections \ref{SecMathQuanTele} and \ref{SecFirstTransformation}, and may introduce Quantum Port in the following sequence:
\begin{enumerate}
\item The moderator begins by providing a motivation to the game. It can be in the form of storytelling so that the player will be invested during the gaming session. We often begin with the story of Alice and Bob who share a special connection that allows for some unconventional communication method.
\item The moderator first introduces the \texttt{Circuit Cards} and noting that there are two types of \texttt{Circuits}, i.e. the thin \& white wires, and the thick \& black wires. From here onwards, we will use the color of the \texttt{Circuits} for our explanation.
\item The moderator states the conditions on how to build a complete circuit. Specifically, the moderator informs the player that the black wires need to begin with \texttt{Entangler}, followed by at least one black wire, and end with \texttt{Pauli Correction}. Meanwhile, the white wires need to follow the exact number of cards that are used to complete the black wires.
\item Next, the moderator introduces the concept of \texttt{Moves} and \texttt{Turns}, and cover \texttt{Move} 1-3 in Section \ref{subsecTurnandMove}.
\item Afterward, the moderator proceeds to explain \texttt{Move} 5-6 in Section \ref{subsecTurnandMove} as the first way to win the game. The moderator will show the player how to use the \texttt{Dice Rule} to determine the correct \texttt{CDC}.
\item Then, the moderator introduces \texttt{Action Cards}. Special attention should be given to \texttt{Information Revealer} and \texttt{Eavesdropper} so that the player understands why these two cards should be used together.
\item Lastly, the moderator may summarize the whole gaming setup and begin with the \texttt{Beginner} setup of Quantum Port. After the player has played the \texttt{Beginner} setup, the moderator may proceed with the \texttt{Default} setup.
\end{enumerate}

\subsection{Post-game debrief}

After a few rounds of gaming sessions, a debrief session should be conducted to inform the player about the real science involved. The common practice is to begin with a few guiding questions, for example:
\begin{enumerate}
\item What is so special about the black wires?
\item What will happen if you only build the white wires to send the \textit{data}?
\item Will you be able to guess the \texttt{QDC} if you do not know about the outcome of the die roll?
\item What is the probability that you may guess the \texttt{QDC} correctly without knowing the outcome of the die roll, or the face value of \texttt{CDC} with \texttt{Information Revealer}?
\item In reality, will you be able to access to the \texttt{QDC}?
\end{enumerate}

Depending on the technical background of the player, the debrief may go into either a diagrammatic explanation of Section \ref{SecFirstTransformation}, or a more detailed mathematical explanation of Section \ref{SecMathQuanTele}, or both. This is the flexibility of Quantum Port that we have designed to allow different audiences to understand quantum teleportation at their own pace.

The moderator needs to establish the connections between the gaming rules and quantum teleportation through the debrief session. Some of the key points that should be included in the debrief session are as follows:
\begin{enumerate}
\item The black wires represent quantum processes, while the white wires represent classical processes.
\item Classical communication is required in quantum teleportation so that faster-than-light communication will not occur.
\item After Alice performed a quantum measurement, Bob needs to perform Pauli correction on his qubit to recover the information Alice wanted to send. Hence, the last card of the black lines is \texttt{Pauli Correction}.
\item To an eavesdropper, the classical information sent by Alice is not useful information. This shows that quantum teleportation offers communication security by quantum physics, rather than by encryption.
\end{enumerate}

The moderator should highlight to the player that some of the gaming mechanics in Quantum Port is not translatable to the physical reality of quantum teleportation:
\begin{enumerate}
\item An eavesdropper has no access to the entangled qubits shared by Alice and Bob. The \texttt{QDC} serves as a winning indicator rather than a direct relationship to the physical process of quantum teleportation.
\item The qubits shared between Alice and Bob are entangled diagrammatically through the cup and cap in Figure \ref{qt-complete}. The \texttt{Entangler} is supposed to be Alice's state, not the entangling effect between Alice and Bob.
\item The \texttt{Action Cards} are mostly part of the gaming mechanics.
\end{enumerate}

\section{Conclusion} \label{SecConclusion}

In this work, we proposed the gamification of categorical quantum mechanics (CQM) on quantum teleportation as a science game for public engagement, with the aim to introduce quantum teleportation protocol through gaming without utilizing complicated mathematical concepts. For future work, one may conduct an empirical study to compare the effectiveness of using Quantum Port as a teaching and learning tool. Since we employ a consistent diagrammatic framework for quantum theory in Quantum Port, we expect future implementations of other quantum processes can be done by adding new cards as an expansion pack, while also maintaining essential mathematical relationships to flexibly support varying levels of rigor from different backgrounds.

\backmatter
\newpage
\bmhead{Supplementary information}

\section*{Declarations}
\subsection*{Availability of data and materials}

Data sharing is not applicable to this article as no datasets were generated or analysed during the current study. All printable materials related to Quantum Port are available through the Supplementary Materials. For a complete set, print one copy of ``Game rules.pdf", one copy of ``Quantum Port Coins.pdf", two copies of ``Quantum Port Cards.pdf", and prepare a four-sided die.

\subsection*{Competing interests}

The authors declare that they have no competing interests.

\subsection*{Funding}

The prototype of Quantum Port was completely self-funded.

\subsection*{Authors' contributions}

PSC and AR gamified the diagrammatic rules of quantum teleportation into Quantum Port. ANA created the artworks for Quantum Port. PSC, AR and NMS contributed in writing and proofreading the manuscript. All authors read and approved the final manuscript.

\subsection*{Acknowledgements}

Quantum Port was tested in three artscience events:
\begin{enumerate}
\item \textit{Not Art, Not Quantum} co-organized by Clarissa Ai Ling Lee, Andrew Jia Cherng Chong, Pak Shen Choong, and Afiqa Nik Aimi on 16 November 2024;
\item \textit{1st Quantum Valentine Fest} co-organized by Taechasith Kangkhuntod from CreativeLab and The Institute of Fundamental Study, Naresuan University, Thailand on 14 February 2025.
\item \textit{ASEAN Quantum Summit 2025} organized and hosted by Universiti Teknologi Malaysia, Malaysia from 10-12 December 2025.
\end{enumerate}

\newpage
\bibliography{bibliography}

\end{document}